\documentclass{pasa}%

\usepackage{graphicx}

\title[The Views from Bell's Spaceships]{Bell's Spaceships: The Views from Bow and Stern}

\author[Lewis et al.]{Geraint F. Lewis\thanks{\tt geraint.lewis@sydney.edu.au}, 
Luke A. Barnes and 
Martin J. Sticka
\affil{Sydney Institute for Astronomy, School of Physics, A28, The University of Sydney, NSW 2006, Australia}%
}%

\jid{PASA}
\doi{10.1017/pas.\the\year.xxx}
\jyear{\the\year}

\usepackage{aas_macros}
\usepackage{hyperref} 
\hypersetup{colorlinks,citecolor=blue,linkcolor=blue,urlcolor=blue}

\hypersetup{draft}

\begin{document}

\begin{frontmatter}
\maketitle

\begin{abstract}
Unravelling apparent paradoxes has proven to be a powerful tool for understanding the complexities of special relativity.
In this paper, we focus upon one such paradox, namely
Bell's spaceship paradox, examining the relative motion of two uniformly accelerating spaceships.
We consider the view from either spaceship, with the exchange of photons between the two. 
This recovers the well known result that the leading spaceship loses sight of the trailing spaceship as it is redshifted and disappears behind what is known as the `Rindler horizon'. An immediate impact of this is that if either spaceship tries to measure the separation through `radar ranging', bouncing photons off one another, they would both eventually fail to receive any of the photon `pings' that they emit. 
We find that the view from this trailing spaceship is, however, starkly different, initially, seeing the leading spaceship with an increasing blueshift, followed by a decreasing blueshift.
We conclude that, while the leading spaceship loses sight of the trailing spaceship, 
for the trailing spaceship the view of the separation between the two spaceships, and 
the apparent angular size of the leading spaceship,  approach asymptotic values. 
Intriguingly, for particular parametrization of the journey of the two spaceships, these asymptotic values are identical to those properties seen before the spaceships began accelerating, and  the view from the trailing spaceship becomes identical to when the two spaceships were initially at rest.
\end{abstract}

\end{frontmatter}

\section{Introduction}\label{introduction} 
The discussion of apparent paradoxes in physical theories have proven to be powerful means of elucidating key theoretical concepts, and Einstein's relativity is no 
exception. While there has been significant focus upon the `twin paradox' and `barn-and-pole paradox', both occupying a significant number of pages in 
the literature and textbooks, more recent discussions have considered uniformly accelerating spaceships in what has become known as 
`Bell's spaceship paradox' \citep{1959AmJPh..27..517D,Bell:111272}.

In this paper, we re-examine the question of uniform acceleration in special relativity, with a focus upon the view from the two spaceships in Bell's paradox. We consider the exchange of light signals during their flights, recovering established results, as well as  uncovering  novel and intriguing outcomes in terms of the view from the  spaceships.   
The layout of this paper is as follows; in Section~\ref{paradox} we review in detail the discussion of Bell's spaceship paradox
in the literature, while in Section~\ref{approach} we outline the approach adopted in this paper. The results and discussion are presented in Section~\ref{results}, and the paper concludes in Section~\ref{conclusions}.

\section{Bell's Spaceship Paradox}\label{paradox}
Bell's spaceship paradox has a long, and sometimes contradictory, history in the story of special relativity; see 
\cite{
1972AmJPh..40.1170E,
doi:10.1119/1.19161,
2000physics...4024N,
0295-5075-71-5-699,
doi:10.1119/1.2733691,
0143-0807-29-3-N02,
0143-0807-31-2-006,
Franklin2013,
2014PhRvA..89f2103M,
2014SerAJ.188...55R} 
for several examples of contributions to the literature, and the reader is directed to the recent review by  \citet{Flores2011} for a more complete discussion.
The starting point is to consider two spaceships initially at rest with respect to each other.
At a time that is synchronous in the rest frame of the spaceships, they fire their engines and undergo uniform acceleration; here, uniform implies that the crews of the spaceships experience identical and constant `g-forces' as the spaceships accelerate. Given the 
identical nature of the acceleration experienced by the spaceships, their separation in the original rest frame remains constant. Their world-lines in the coordinates of this frame differ only by an offset: $x_\textrm{leading}(t) = x_\textrm{trailing}(t) + L$.

The paradoxical aspect of the Bell's scenario comes from considering a taut thread strung between the two spaceships while they are initially sitting at rest. Once the spaceships begin 
to accelerate, what happens to the thread? There are three options, namely that it remains taut, sags, or stretches until it eventually breaks. A cursory examination of the problem might suggest that, due to relativistic length contraction, the distance between the two spaceships decreases and so the thread will sag. Or, the string and the distance between the spaceships will contract equally, so the thing will be unaffected. However, when examined in detail, it is found that the distance between the spaceships increases, so the string snaps.

Perhaps the clearest way to see that the string snaps is presented by \citet{maudlinspace-time}. Suppose that after some time, $t_\textrm{end}$, both spaceships shut off their engines and return to inertial motion at coordinate velocity $v_\textrm{end}$ in the original rest frame. Note that this can be achieved by the spaceship pilots agreeing to burn their engines for exactly the same amount of proper time before shutting them down. While the rockets shutting down their engines is seen to be simultaneous in the original rest frame, these events will not be simultaneous to the pilots on-board the spaceships \citep[e.g.][]{1989AmJPh..57..791B}.  
However, the resultant world-lines after the acceleration are straight and parallel. If the string returns to the same equilibrium state as before the acceleration, with the same forces balanced between atoms that make up the string, then a beam of light bounced off the far end of the string will return in the same amount of time as before the acceleration. If the light takes longer to return, then this can only mean that the distance is greater and the string must have stretched and snapped; we will return to this question of `radar-ranging' through the exchange of photons in detail in Section~\ref{radar}, but here we give an outline of the results.

\begin{figure}
\begin{center}
\includegraphics[width=\columnwidth]{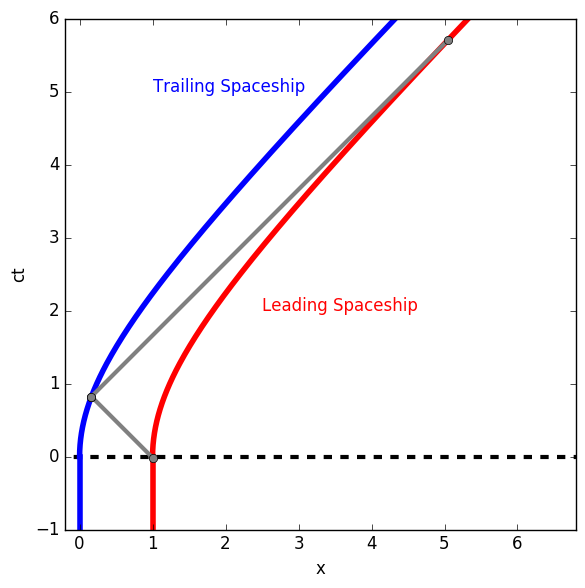}
\caption{An illustrative space-time diagram of the paths of the two spaceships under consideration for this paper, with the leading spaceship  shown in red, whereas the trailing spaceship  shown in blue. Example exchanges of photons are shown in grey.} 
\label{paths} 
\end{center}
\end{figure}

Figure \ref{paths} presents a systematic representation of two spaceships under consideration in Bell's paradox, with the red line indicating the leading spaceship, while the blue line is the trailing spaceship. The grey line denotes the  the path of a  light ray exchanged between the leading and trailing spaceship. 
Before the acceleration, the bouncing light returns to the stationary spaceship after $\Delta t_\textrm{before} = 2$. We can similarly trace a radar-ranging photon on a space-time diagram after the acceleration has ceased and the spaceships are in uniform motion, and calculate the proper time between its emission and reception. We find that the time taken for the photon to return is $\Delta t_\textrm{after} = 2 / \sqrt{1 - v^2_\textrm{end}/c^2}$. The distance between the spaceships has expanded, and the string has snapped.

We can also reconstruct the series of events in the instantaneous reference frames of each of the spaceships. Consider the trailing spaceship at some time during the accelerating phase. In the coordinates of the original rest frame, the simultaneity slice of the rocket is tilted upwards with respect to the $x$-axis.
Given an event at time t on the world-line of the trailing rocket, the simultaneous event (according to the trailing rocket) on the world-line of the leading rocket is at a larger value of $t$.
 While the acceleration is symmetric in the original frame, the trailing rocket concludes (having reconstructed the series of events in its instantaneous frame) that the leading ship is in fact accelerating more rapidly, as it is travelling faster at any given time. Conversely, the leading ship concludes that the trailing ship accelerated more slowly; thus, the string snaps. This is not merely a consequence of the finite speed of light. That is, it is not about what the trailing ship \emph{sees}. It is consequence of the relativity of simultaneity, and the changing slices of simultaneity as the spaceships accelerate. 

The majority of the literature on Bell's spaceship paradox has focused upon this question of what happens to the taut string between the ship, with examinations of the question of relativistic stress in accelerating objects.
However, in the remainder of this paper, we will turn to less-explored questions regarding the relativistic influence on their view of each other.

\section{Approach}\label{approach}
To understand the view from each spaceship, we will  consider  acceleration within special relativity,  a topic which has been discussed in  detail elsewhere
[see Chapter 5 of general relativity by \citet{hartle2003gravity} for an excellent discussion] and here we provide a summary of the key points. The motion of an accelerating spaceship through space-time is described by a 4-velocity, $u^\alpha$, and 4-acceleration, $a^\alpha$, the motion obeys key normalization relationships, namely;
$$\setlength\arraycolsep{0.4em}
\begin{array}{ccccc}
 u \cdot u &=& \eta_{\alpha\beta}\ u^\alpha\ u^\beta &=& -c^2 \nonumber \\
 a \cdot a &=& \eta_{\alpha\beta}\ a^\alpha\ a^\beta &=& g^2 \nonumber \\
 u \cdot a &=& \eta_{\alpha\beta}\ u^\alpha\ a^\beta &=& 0 
\end{array}
$$
where $c$ is the speed of light.
Here, $\eta_{\alpha\beta}$ are the components of the Minkowski metric of flat space-time, and
$g$ is the magnitude of the acceleration, experienced as a g-force within the spaceship. The final expression here demonstrates that the 4-velocity and 4-acceleration remain orthogonal during the motion; see the work of  \citet{1960PhRv..119.2082R} for a detailed discussion of orthogonality and 
the hyperbolic nature of relativistic motion.

In this paper, we will consider uniform acceleration, such that the magnitude of the acceleration is the same in the instantaneous reference frame aligned with the spaceship. For motion purely in the $x-$direction, the world-line of a spaceship has an analytic form, namely;
\begin{align}
x^\alpha & = \left( c t(\tau) , x( \tau ) \right) \nonumber \\
 & = \frac{c^2}{g} \left( \sinh\left(\frac{g \tau}{c}\right) , \cosh\left(\frac{g \tau}{c} \right) + C \right) 
 \label{analyticpaths}
\end{align}
where $\tau$ is the proper time experienced by an observer on the accelerating spaceship.  The
acceleration begins at the proper time of $\tau=0$, corresponding to a coordinate time of $t=0$, and  $C$ is a constant set by the starting spatial location of the spaceship. The components of the 4-velocity are given by
\begin{equation}
u^\alpha = \left( c \frac{dt}{d\tau} , \frac{dx}{d\tau}\right) = c 
\left( \cosh\left(\frac{g \tau}{c}\right) , \sinh\left(\frac{g \tau}{c} \right) \right)
\end{equation}

In examining the situation as described in Bell's spaceship paradox, the two spaceships are initially considered to be at rest within a particular coordinate system, with one spaceship at the origin, $x=0$, while another is located at $x=L$; clearly, in this reference frame, the two spaceships are separated by $L$. At a coordinate time of $t=0$, the two spaceships fire their engines to produce an identical uniform acceleration, and the motion  is described by the above equations. The space-time diagram shown in Figure~\ref{paths} demonstrates these paths, with the leading spaceship shown in red, while the trailing spaceship is shown in blue. Again, note that the separation between the two spaceships in the initial reference frame remains $L$ for the duration of the journey.

Within the flat space-time of special relativity described by the Minkowski metric, the null paths traced by photons are lines at $45^o$ in $(ct,x)$ coordinates, and therefore the trajectories of photons exchanged between the two spaceships can be determined from purely geometric considerations; two example paths of photons that are exchanged between the two spaceships are shown in Figure~\ref{paths}.
To determine the relative energies of the exchanged photons, we can use the fact that the energy of a photon with a null 4-momentum of $k^\alpha = (k^t , k^x)$ as seen by an observer with a 4-velocity of $u^\alpha = (u^t , u^x)$ is given by
\begin{equation}
E = - k \cdot u = -\eta_{\alpha\beta} k^\alpha u^\beta
\end{equation}
[see \citet{1994AmJPh..62..903N} for a generalised discussions of redshifts in relativity].
With this, the energy of a photon as observed by one spaceship, $E_o$, compared to the energy it was emitted by the other spaceship, $E_e$, is given by
\begin{equation}
\frac{E_o}{E_e} = \frac{ \eta_{\alpha\beta} k^\alpha u_o^\beta }{\eta_{\alpha\beta} k^\alpha u_e^\beta}
\end{equation}
Given that photons follow null paths, in the Minkowski space-time $k^t = | k^x |$, where the absolute aspect of the spatial component accounts for
photon paths in either the positive or negative $x-$direction. With this, the relative energies of the photons is given by
\begin{equation}
\frac{E_o}{E_e} = \frac{u^t_o \mp u^x_o}{u^t_e \mp u^x_e}
\end{equation}
where the positive term is for photons travelling in the negative $x-$direction, while the negative term is for photons travelling in the positive $x-$direction.

\section{Results}\label{results}
Given the analytic form of the 4-velocity in Equation~\ref{analyticpaths}, and the geometric form of the photon path, we are able to drive expressions for the relative emitted and observed photon energies between the two spaceships. In the following, the subscript, $l$, refers to the leading spaceship, while $t$ is the trailing spaceship. Additionally, we adopt units in which the speed of light, $c=1$, in which we employ a normalized acceleration given by $a = \frac{g}{c}$.

\subsection{Radar Distance}\label{radar}
\begin{figure}
\begin{center}\includegraphics[width=\columnwidth]{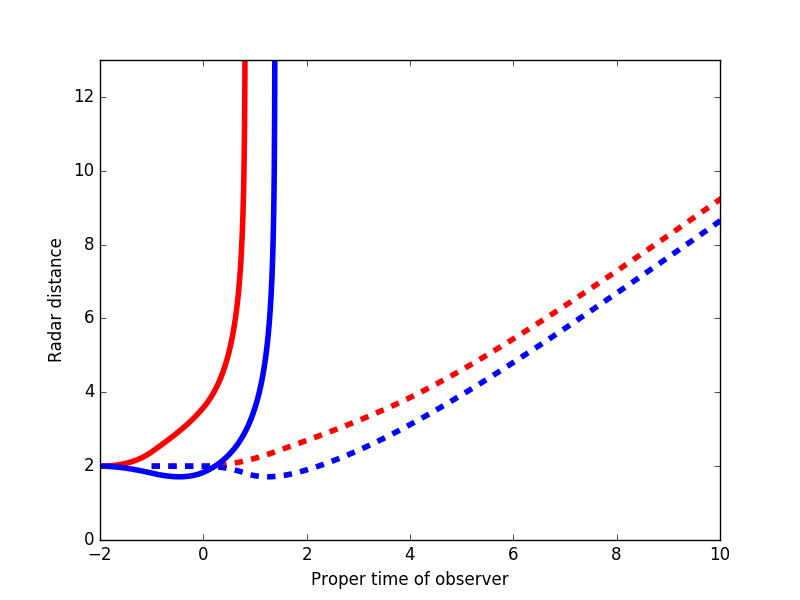}
\caption{The radar distance as measured by the leading spaceship  in red as a function of the proper time photons are emitted (solid), or subsequently received (dashed). The corresponding radar distance for the trailing spaceship  is presented in blue. As for the previous figures, this is presented for a fiducial case where $a=0.5$ and $L=1$.} 
\label{radardistance} 
\end{center}
\end{figure}

The concept of radar distance is well established in relativistic physics, with an observer measuring the time taken for a photon to travel to an object of interest and back again. In Figure~\ref{paths}, the grey photon path from leading spaceship to trailing spaceship and back again to the leading spaceship, representing a single `ping' in the leading spaceship's determination of the radar distance to the trailing spaceship \citep{2008MNRAS.388..960L,2008ASSL..349..131P}.
Given the analytic form of the spaceship paths given by Equation~\ref{analyticpaths} we can relate the proper times for the emission and receipt of light rays during the accelerated portion of the spaceships' journeys. For a light ray travelling in the positive $x-$direction, this is given by
\begin{equation}
\exp( -a \tau_l ) = \exp( -a \tau_t ) - a\ L
\end{equation}
whereas for a light ray travelling in the negative $x-$direction, the corresponding expression is 
\begin{equation}
\exp( a \tau_l ) = \exp( a \tau_t ) - a\ L
\end{equation}
Using these expressions, it is straight-forward to calculate the radar distance determined by the two spaceships and these are presented in Figure~\ref{radardistance}, for an acceleration of $a=0.5$ and $L=1$; we note that there is a degeneracy between the separation and acceleration, with present equations above dependent upon only their (dimensionless) product, namely $a\ L$.  
Here, the solid line denoted the light travel time as a function of the proper time that a photon is emitted, whereas the dashed line is as a function of the proper time that a photon is received. The red curves represent the view from the leading spaceship, whereas the blue curves are for the trailing spaceships.

In examining the solid curves in Figure~\ref{radardistance} it is immediately apparent that the radar distance for both spaceships diverges at a finite proper time. This is understandable due to presence of the Rindler horizon for the leading spaceship as eventually it will outrun any photon emitted or reflected from the trailing spaceship, and so the return journey required for the radar `pings' become impossible. But as we have seen previously, while the leading spaceship loses sight of the trailing spaceship as it slips behind the Rindler horizon, the trailing ship continues to view the leading spaceship, even though it can no longer measure a finite radar distance. So it would be able to see the leading ship's tail-lights, but would be unable to illuminate the ship with its own headlights. This also confirms the resolution of the original Bell's paradox: as the ships accelerate, the (radar-ranging) distance between them increases, snapping the string. In the final section of this paper, we will consider the apparent angular size of the two spaceships as determined by those travelling on them.

\subsection{Blueshift and Redshift}

\subsubsection{View from the leading spaceship:}\label{leading}
When considering the view from the leading spaceship, it must be remembered that when it begins its acceleration, it is still receiving photons from the trailing ship that were emitted before it begins its acceleration. So, for the initial stages of its journey, the 4-velocity for the photon emitter is given by $u_e^\alpha = (u_e^t,u_e^x) = (1,0)$. Eventually, the leading spaceship will receive photons from the trailing spaceship after it begins its acceleration, where the 4-velocity is given by the above relations. 

With this, for the first stage of the journey, the relative energies of the photons is given by
\begin{equation}
\frac{E_l}{E_t} = \exp( { - a \tau_l } )
\end{equation}
while the second stage of the path, this is given by
\begin{equation}
\frac{E_l}{E_t} = \frac{1}{1 + a\ L\ \exp({a \tau_l})}
\label{eq1}
\end{equation}
where $\tau_1$ is the proper time recorded on a clock on the leading spaceship.
The transition between these two relations occurs at 
\begin{equation}
\tau_l = -\frac{1}{a} \log_e \left( 1 - a\ L \right)
\end{equation}

\subsubsection{View from the trailing spaceship:}\label{trailing}
As with the view of the leading spaceship, when the trailing spaceship begins its acceleration it is still receiving photons from the leading spaceship when it was at rest, so again $u_e^\alpha = (u_e^t,u_e^x) = (1,0)$, and, again, the trailing spaceship begins to receive photons from the leading spaceship once it begins its acceleration. In this circumstance, the ratio of the photon energies for the first stage of the accelerated path is given by 
\begin{equation}
\frac{E_t}{E_l} = \exp( { a \tau_t } )
\end{equation}
and the corresponding values for the second part of the journey is given by
\begin{equation}
\frac{E_t}{E_l} = \frac{1}{1 - a\ L\ \exp({- a \tau_t})}
\label{eq2}
\end{equation}
where $\tau_2$ is the proper time recorded on a clock on the trailing spaceship.
The transition between these two relations occurs at 
\begin{equation}
\tau_t = \frac{1}{a} \log_e \left( 1 + a\ L \right)
\end{equation}

\begin{figure}
\begin{center}
\includegraphics[width=1.0\columnwidth]{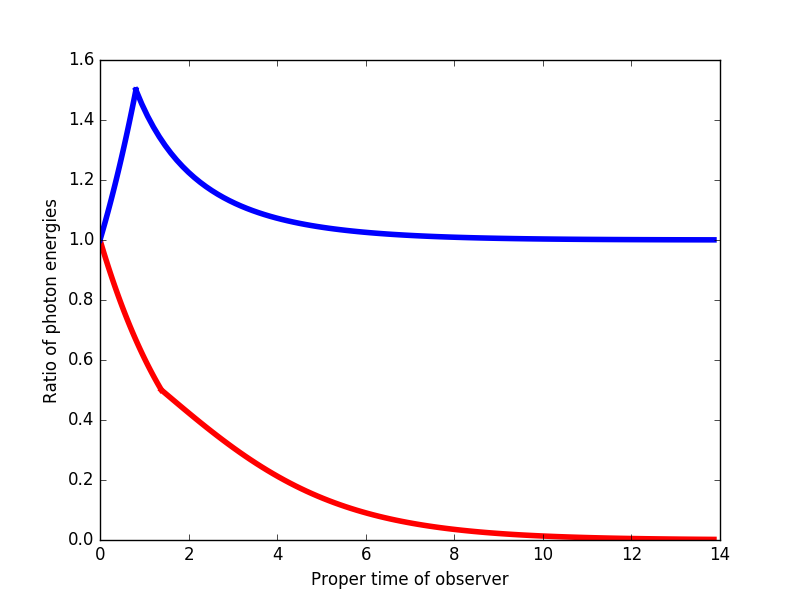}
\caption{The observed blue/redshift of photons as seen by the leading spaceship (red curve) and trailing spaceship (blue curve) as a function of the proper time of the observe. The distinct breaks in each of the curves delineates the point in the journey where there is a transition between observing the emitting spaceship from being stationary to accelerating. As noted in the text, the leading spaceship loses sight of the trailing ship, with the observed energy tending to zero. The trailing spaceship initially sees an increase in the blueshifting of the leading spaceship, before it decreases back towards unity.} 
\label{redshift} 
\end{center}
\end{figure}

\subsubsection{Interpretation:}\label{interpretation}
Figure~\ref{redshift} presents a graphical illustration of these results, again for an initial separation of $L=1.0$ and an acceleration of $a=0.5$, with the red curve representing the view from the leading spaceship, while the blue curve is the view from the trailing ship; again, we note that the results depend only upon the product of these two quantities.
As per the above expressions, the leading spaceship's views the trailing ship as being increasingly redshifted, with the observed energy dropping to zero as the image of the trailing ship is frozen on what is known as the Rindler horizon \citep{1966AmJPh..34.1174R}. 

\begin{figure}
\begin{center}
\includegraphics[width=\columnwidth]{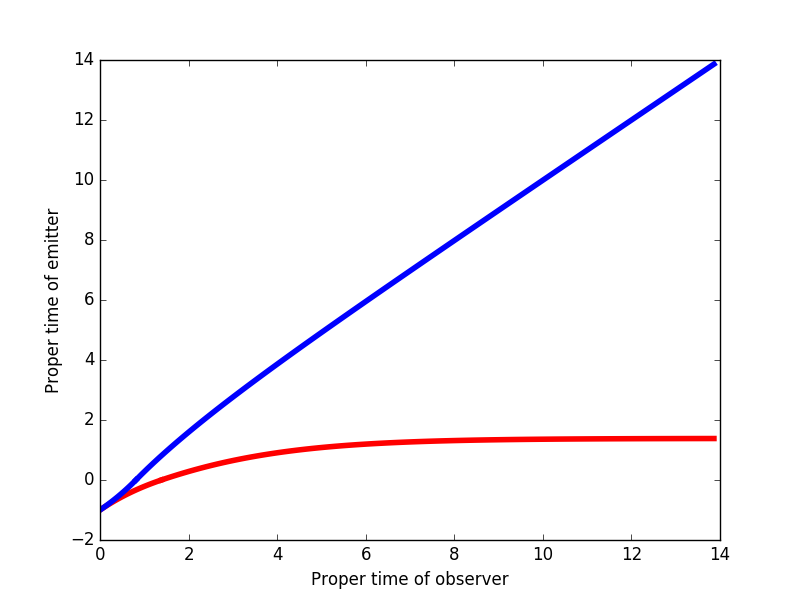}
\caption{The relationship between the proper time of the emitting spaceship when a photon is emitted, compared to the proper time on the observing spaceship when the photon is received. The red curve represents the case of a photon being emitted from the trailing spaceship and being observed by the leading spaceship, while the blue curve corresponds to the emission of photons from the leading spaceship and being observed by the trailing spaceship. } 
\label{time} 
\end{center}
\end{figure}

This freezing on the Rindler horizon can also be seen in Figure~\ref{time} which presents the proper time on the emitting spaceship when a photon is emitted, and the corresponding time on the receiving spaceship when the photon is observed. Again, the red curve corresponds to the case where photons are emitted from the trailing spaceship and observed by the leading spaceship, whereas the blue curve is the case where photons are emitted from the leading spaceship and received by the trailing spaceship. Considering the red curve, it is clear that proper time of the emitter is asymptoting to a particular value as the redshift increases, demonstrating the freezing of the image of the trailing spaceship on the Rindler horizon.

Examining the blue curves in Figures~\ref{redshift} and \ref{time} reveals that the view from trailing spaceship is quite different, with the asymptotic behaviour of the relative energies tending to unity, so there is no net redshifting or blueshifting, and the relative ticking of the clocks becoming in sync. 
Hence, the trailing ship appears to settle back into the view before the acceleration began.
We will return to the interpret the asymptotic behaviour of the emitted and observed photon energies in the next section.

In re-examining the results presented in equations \ref{eq1} and \ref{eq2}, in the limit where the acceleration and separation as small, we note that the exponential terms tends to unity and the ratio of the observed photon energies depend upon $a\ L$. The form is equivalent of the observed photon redshifting and blueshifting in a uniform weak gravitational field which is dependent upon the potential difference between the emitter and the observer, with $a\ L$ being equivalent to $\Delta \Phi = g h$; such a result is expected from the relativistic equivalence principle of uniform acceleration and a uniform gravitation field, and the reader is directed to Chapter 6 of \citet{hartle2003gravity} for more details. 

\subsection{Angular Size}\label{angular}
Unlike radar ranging, which requires a two-way exchange of light, ``seeing" only requires light to travel in one direction. For the purposes of this study, we will consider the angular size of each spaceship as determined from the other.
In calculating this, we assume each spaceship is a disk of radius $d$ orientated perpendicular to the $x-$direction. It is straight-forward to connect light paths leaving the edge of one of the spaceships to a camera at the centre of the other by using the fact that these paths must be null;
\begin{equation}
-(t(\tau_i) - t(\tau_j))^2 + (x(\tau_i) - x(\tau_j))^2 + d^2 = 0
\label{null}
\end{equation}
where $\tau_i$ and $\tau_j$ are the proper times experienced on each of the spaceships. For the purposes of this study, we decided to employ a numerical root-finder to identify the null paths.

Once the light paths are identified in space-time, their orientation relative to the camera can be determined. However, it is essential to transform these into the observer's frame to determine the angle as discerned by the camera. For this, we identify an orthonormal frame (see \citet{hartle2003gravity}, chapter 8 for details) with the observer and can transform the $x-$component of the photon 4-momentum to be
\begin{equation}
p'^x = p^x u^t - p^t u^x
\end{equation}
where $u^t$ and $u^x$ are the components of the 4-velocity of the observer, and $p^t$ is determined from the fact that photon paths are null. 

\begin{figure}
\begin{center}
\includegraphics[width=\columnwidth]{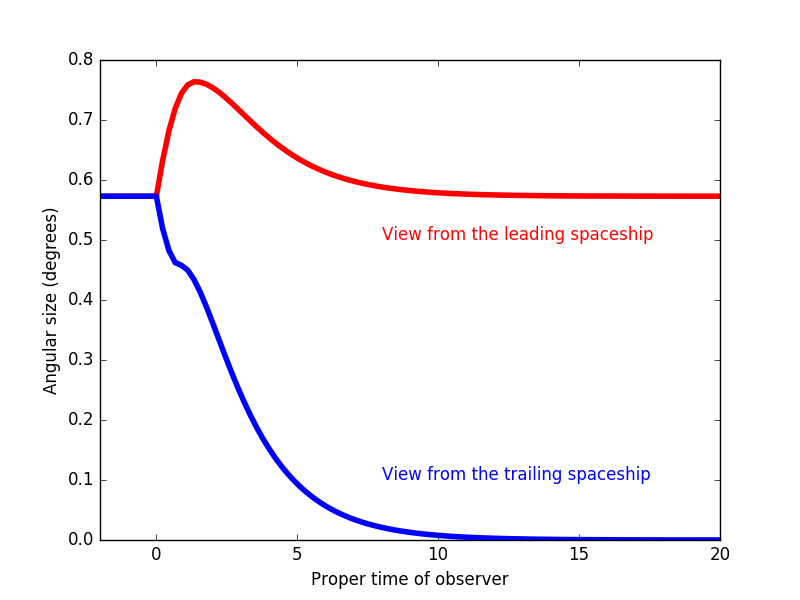}
\caption{The apparent angular size of the two spaceships as viewed from each other, with the red being the view from the leading spaceship, and the blue being the view from the trailing spaceship. As in previous examples, $L=1$, $a=0.5$ and the radius of the spaceships is $d=0.01$. } 
\label{angularfig} 
\end{center}
\end{figure}

For the purposes of this study, we solve this problem numerically and in Figure~\ref{angularfig} we present the apparent angular size of the two spaceships. For this, we assume that $d=0.01$, with $L=1$ and $a=0.5$ assumed previously, with the red curve denoting the view from the leading spaceship, while the blue curve represents the view from the trailing spaceship. 

The view from each spaceship is starkly different. The trailing spaceship sees the leading steadily decreases in angular size, getting smaller and smaller as the two spaceships accelerate. However, the leading spaceship initially sees the trailing spaceship grow in size; this, of course, is a well-known special relativistic result, where an observer at relativistic speeds sees objects, such as a distant star-field, apparently pile up in the direction of motion, a relativistic aberration effect\footnote{e.g. {\tt math.ucr.edu/home/baez/physics/Relativity... /SR/Spaceship/spaceship.html}}. 
This is reflected in the linear change in the apparent angular size of the spaceships once the acceleration starts. In the low velocity limit, we can approximate the velocity to be $v \sim a \tau$ and the apparent angular size given by special relativistic aberration relation is given by
\begin{equation}
\frac{\theta'}{\theta} \sim \left( 1 \pm v \right) \sim \left( 1 \pm a \tau \right)
\end{equation}
accurately describing this initial linear behaviour at the start of the acceleration.
But, as the acceleration continues, the angular size deviates from the linear relationship, and the view seen by the leading spaceship decreases, tending back to the angular size seen before the acceleration starts. 

\begin{figure}
\begin{center}\includegraphics[width=\columnwidth]{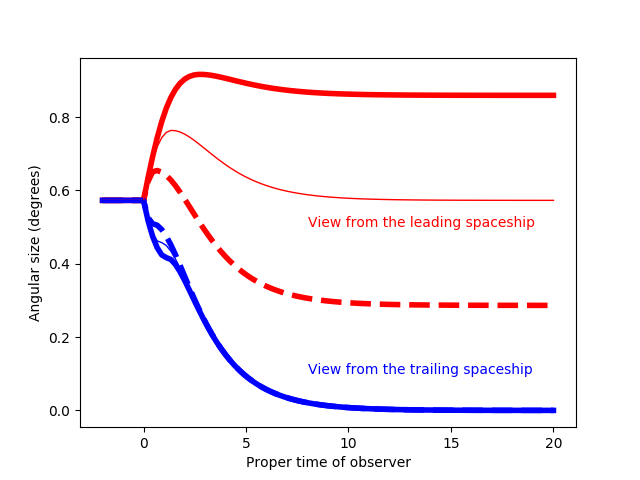}
\caption{As Figure~\ref{angularfig}, with an acceleration of $a=0.5$, but different values of the initial separation ($L$) and angular size ($d$). Again, the red lines corresponds to the view from the leading spaceship, while the blue is the view from the trailing spaceship. The thin line represents the situation where $L=1$ and $d=0.01$ (as in the previous figure), while the thicker solid line corresponds to $L=0.5$ and $d=0.005$, and the thicker dashed line is for $L=1.5$ and $d=0.015$. In each case, the the apparent angular size before the period of acceleration is the same, but it is clear that the asymptotic angular size of the trailing spaceship as seen by the leading spaceship depends upon the chosen values of $L$ and $d$.} 
\label{angularfig2} 
\end{center}
\end{figure}

An exploration of the magnitude of the acceleration and the separation of the spaceships reveal these views to be generic, with the trailing spaceship seeing a diminishing size for the leading spaceship, whereas the the leading spaceship sees the angular size of the trailing ship asymptote to some fixed value. This is generally not the size of the spaceship as seen before the acceleration starts, and the particular asymptote see in Figure~\ref{angularfig} is due to the particular choice of $a$ and $L$. Will illustrate this in Figure~\ref{angularfig2}, which presents the case in the previous figure, but also adding the case where $L=0.5$ and $d=0.005$ (thick solid line) and $L=1.5$ and $d=0.015$ (thick dashed line). With these, the apparent size of the spaceships before the acceleration is the same, and after the acceleration, the trailing ship sees the size of the leading spaceship continue to decrease, whereas the leading spaceship sees the size of the trailing spaceship asymptote to a constant value, although the precise value of this asymptote depends explicitly of the chosen values of $L$ and $d$.

\section{Conclusions}\label{conclusions}
In this paper, we have revisited Bell's spaceship paradox, expanding the problem to consider the view from both the leading and trailing spaceships, with the exchange of photons between the two during their journeys, a previously unexplored aspect of the  scenario of identically accelerating spaceships.

As well as recovering established results in Bell's spaceships, 
we also find that, due to the presence of the Rindler horizon, the distance between the two spaceships as determined by radar ranging diverges for both spaceships. However, the view from each spaceship  does not reflect this divergence and are distinctly different.

For the leading spaceship, we find that it sees the trailing spaceship being progressively redshifted and shrinking as it vanishes behind its Rindler horizon. The view from the trailing spaceship is quite different, finding that there is an initial increase in the blueshifting of photons, before a subsequent decrease. Similarly, the angular size of the two spaceships has some strange behaviour, with the trailing spaceship seeing the leading spaceship shrink in size, whereas the leading spaceship sees the angular size of the trailing spaceship asymptote to some particular value.  
 This behaviour is quite peculiar and unexpected, and, in conclusion, while Bell's spaceship paradox is well studied and discussed in the field of relativity, it still has some surprises yet to yield.

\section*{Acknowledgements} 
We thank the referee for their positive comments and additional relativistic insights that improved the presentation of this paper. 
LAB is supported by a grant from the John Templeton Foundation. This publication was made possible through the support of a grant from the John Templeton Foundation. The opinions expressed in this publication are those of the author and do not necessarily reflect the views of the John Templeton Foundation.

\newpage
\bibliographystyle{pasa-mnras}
\bibliography{references}

\end{document}